\documentstyle[aps,epsfig,12pt,tighten,floats]{revtex}
\begin{document}
\newcommand{\identity}{\:\mbox{\sf 1} \hspace{-0.37em} \mbox{\sf 1}\,}
\draft

\title{Generation of entangled states and error protection from 
adiabatic avoided level crossings}
\author{Nicole F. Bell$^1$\footnote{Present address: 
NASA/Fermilab Astrophysics Center, Fermi National Accelerator Laboratory,
Batavia, Illinois 60510-0500}, 
R. F. Sawyer$^2$, 
Raymond R. Volkas$^1$ and
Yvonne Y. Y. Wong$^1$\footnote{Present address:
Department of Physics and Astronomy, University of Delaware, Newark, 
Delaware 19716}
}

\address{\bigskip $^1$%
School of Physics, Research Centre for High Energy Physics\\ The
University of Melbourne, Victoria 3010, Australia\\ \smallskip $^2$%
Department of Physics, University of California at Santa Barbara
\\ Santa Barbara, California 93106\\ \medskip
(n.bell@physics.unimelb.edu.au, sawyer@vulcan.physics.ucsb.edu,
r.volkas@physics.unimelb.edu.au, y.wong@physics.unimelb.edu.au)}
\maketitle

\bigskip
\bigskip

\begin{abstract}

We consider the environment-affected
dynamics of $N$ self-interacting particles living in one-dimensional 
double wells. Two topics are dealt with. First,
we consider the production of entangled states of two-level systems.
We show that by adiabatically varying the well biases we may dynamically 
generate maximally entangled states, starting from initially 
unentangled product states. Entanglement degradation due
to a common type of environmental influence is 
then computed by solving a master
equation. However, we also demonstrate that entanglement
production is unaffected if the system-environment coupling
is of the type that induces ``motional narrowing''.
As our second but related topic, we construct a different master equation
that seamlessly merges error protection/detection 
dynamics for quantum information
with the environmental couplings responsible for
producing the errors in the first place.
Adiabatic avoided crossing schemes are used in
both topics.
\end{abstract}

\section{Introduction}

Real quantum systems are always coupled to environments. Research in the
last two or three decades has shown that system--environment interactions
greatly affect quantal coherences in the system. Open-system dynamics can
often be described by a master equation for the reduced density matrix of 
the system. The full Hamiltonian is written as the sum of system-only, 
environment-only and system--environment interaction terms. The master
equation for the system then depends on the system-only Hamiltonian
plus other terms derived from the fundamental system--environment 
couplings. These additional pieces induce an effective non-unitary
evolution for the system that can, for example, lead to an apparent loss
of quantal coherence.

In many studies, the system-only Hamiltonian is taken to be the sum of
one-body terms. An example of relevance to this paper arises for a system of
$N$ particles, each in a one-dimensional symmetric double well.  The system
Hamiltonian
\begin{equation}
H_{\text{sys}} = \sum_{i = 1}^{N} 
\left( E^{(i)} \identity^{\!\!(i)} + \omega^{(i)} \sigma_x^{(i)} \right)
\end{equation}
describes particle $i$ oscillating with frequency $\omega_i$ between the
left and right sides of its well independently of the other particles
in the system. When system--environment coupling is added, this behaviour 
can be drastically modified, with the quantum Zeno \cite{qze} or freezing 
phenomenon being the most extreme example.

The purpose of this paper is to contribute to a study of closed and open quantum
system evolution that goes beyond one-body system Hamiltonians by including
interaction terms between the system particles. This is a vast and extremely
rich field of enquiry, with many possible lines of development. 
For reasons to be explained below we will focus on two topics.
The first is entanglement creation through adiabatic evolution and its
degradation due to environmental influences. The second is quantum
information error protection through adiabatic evolution.

Common to both topics is the use of adiabatic avoided level crossing dynamics. 
This special case of quantum mechanical time evolution
has applications in diverse areas of physics,
from the Mikheyev-Smirnov-Wolfenstein effect for neutrino oscillations \cite{msw}
to quantum computation \cite{farhi}. We use it here because it allows
non-trivial dynamics to occur while retaining both an element of theoretical
simplicity plus visualisability with the aid of level crossing diagrams.

Whenever one has more than a single particle in one's quantum system, 
entanglement adds spice to the analysis.
Entangled states are a unique feature of quantum mechanics, first
studied in connection with the issue of non-locality
\cite{eprbell}. They are also of critical importance in
discussions of decoherence, measurement and the
quantum-to-classical transition \cite{zurek}. More recently,
entanglement has emerged as a useful resource in the field of
quantum information theory, enabling processes such as quantum
cryptography \cite{crypt} and teleportation \cite{teleport}.  The
production and manipulation of entangled states is a key element
in any realisation of a quantum computer \cite{qc}.

As our first topic, we describe an efficient method for the preparation of entangled
states which utilises the level crossing structure of the system's
energy eigenstates.  Specifically, we discuss the Bell states in the 
case of a bipartite system, $W$ states in a tripartite system and 
generalised $W_N$ states in multi-partite systems. We then subject
the evolving systems to environmental interactions
that affect the entanglement production process. We show that the outcome
depends on the form of system-environment coupling: entanglement
degradation is the generic effect, but the special case of
``motional narrowing'' noise actually leads to zero degradation.
Calculations presented in Ref.\cite{bsv} simply took initial Bell states
and computed the evolution of entanglement in the presence of
environmental noise. The present work is a natural extension of this.

As a separate but related topic, we show that
a method of error protection 
using entangled states emerges from a variation of the above procedure.
We devise a scheme to protect the information encoded in a single qubit 
from bitflip and phaseflip errors. Our goal is to construct a
master equation which seamlessly joins the dynamics of qubit encoding,
error protection and decoding to the environment-coupling influences
responsible for error generation. While the outcome of the
open system evolution can be dissected into these quantum information
theoretic subprocesses, the actual dynamics is simply what you get
when you solve a certain master equation. 

The building block for our multi-partite systems is the two-state system
of a particle in a one-dimensional double
well, chosen for its generic properties and broad interest (see
Ref.\cite{leggett}). The well is taken to be partitioned by a
barrier of height $V$ that is large compared to the ground state
energy $E_0$. (The two lowest energy levels will be collectively
called the ``ground state''. For a symmetric well with $V \gg E_0$
they are nearly degenerate.)
Adopting quantum information theoretic parlance, we
denote by ``qubits'' the
states occupying the two-dimensional Hilbert space
spanned by the left and right sides of the well (in
the ground state). The left and
right eigenstates, denoted $| 1\rangle$ and $| 0
\rangle$, correspond
to the $+1$ and $-1$ eigenvalues of $\sigma_z$, respectively.
For an isolated well, the Hamiltonian has the form
\begin{equation}
H_1 = E_0 \identity + \omega \sigma_x + f \sigma_z, \label{1ham}
\end{equation}
where the energy splitting $\omega$ is essentially a
tunneling rate between the two sides of the well, and the
parameter $f$ is the well bias.

\section{Generation of entangled states}
\subsection{Two qubit system}

We begin with a bipartite two-state system, the simplest system in
which one may discuss entanglement. Our aim here is to produce one
of the Bell states,
\begin{equation}
|\Psi^{\pm} \rangle = \frac{1}{\sqrt{2}}(|10 \rangle \pm |01
\rangle),
\end{equation}
dynamically from certain initially factorisable states.

Let us denote the two-particle (reduced) 
density matrix of qubits 1 and 2 as
\begin{equation}
\rho = \rho_1 \otimes \rho_2, 
\end{equation}
and consider a Hamiltonian of the form presented in Eq.\ (\ref{1ham})
for each of the two qubits, together with a third term which
couples them:
\begin{equation}
\label{puffin}
H_2 = \left[ \omega \sigma_{x1} + f(t)\sigma_{z1}\right]
\otimes \identity + \identity \otimes \left[ \omega  \sigma_{x2}
+ f(t) \sigma_{z2} \right] + \lambda \sigma_{z1} \otimes
\sigma_{z2}. \label{2ham}
\end{equation}
Observe that the Hamiltonian is invariant under the interchange 
of the two qubits.  For simplicity, only
the well biases $f$ are allowed to vary with time, and we
demand that $\omega \alt 0.1 \lambda$. This last condition is
necessary in order for the desired level crossing structure to
arise. (It is implicitly assumed that only the lowest
states in each well are populated. We will need to relax this condition 
later on when we couple the system to a hot bath.)

\begin{figure}[ht]
\begin{center}
\epsfig{file=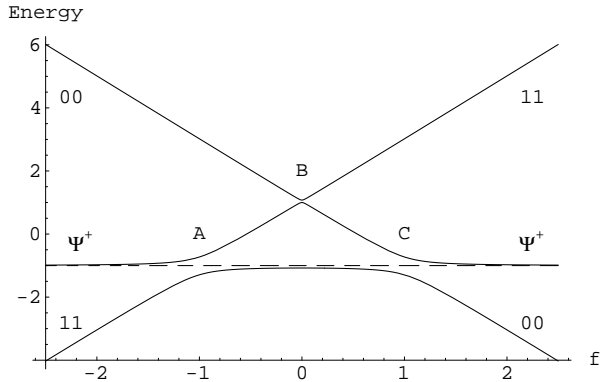,width=8cm} \caption{\label{2qubit} Level
crossing diagram for the two-qubit system. 
The energy and the variable $f$ are in
units of $\lambda$, and the dotted line denotes the decoupled
state $|\Psi^- \rangle$.}
\end{center}
\end{figure}

The four
eigenstates separate into an antisymmetric singlet state
$|\Psi^- \rangle$ for all values of $\omega$, $f$ and $\lambda$,
and a symmetric triplet composed of linear superpositions of the
set ${\cal S} = \{ |11\rangle, |00\rangle, |\Psi^+ \rangle \}$,
which asymptotes to ${\cal S}$ for large $|f|$'s. Note that the
permutation symmetry of the Hamiltonian ensures that the evolution
of the singlet state is disconnected from that of the triplet
states at all times so we are dealing with an effectively
three-state system. In Fig.\ \ref{2qubit}, we plot the energy
level diagram as a function of $f$.

The states $|11\rangle$ and $|00\rangle$ experience avoided level
crossings with  $|\Psi^+ \rangle$ at $f=-\lambda$ and $f=\lambda$
respectively, labelled points $A$ and $C$ in Fig.\ \ref{2qubit}.
Hence, an initial state $|\phi (t=0) \rangle = |11 \rangle$
created at $f \ll -\lambda$ will continue to reside in the lowest
energy eigenstate provided that $f$ varies sufficiently slowly
with time, and be adiabatically transformed, across point $A$,
 to the entangled state
\begin{equation}
\label{marshmallow} |\phi (t=T) \rangle \simeq |\Psi^+ \rangle -
\frac{\omega}{\sqrt{2}(f+\lambda)} |11 \rangle +
\frac{\omega}{\sqrt{2}(f-\lambda)} |00 \rangle + {\cal
O}(\omega^2),
\end{equation}
where $T$ corresponds to any point in the period of time during
which $f$ lies in the range $-\lambda < f < \lambda$. Further
adiabatic evolution beyond point $C$ will turn $|\phi (t) \rangle$
once more into a factorisable state $|00 \rangle$.  Note in Eq.\
(\ref{marshmallow}) that the entangled state $|\phi(t=T) \rangle$
inevitably contains some order $\omega/[f+\lambda,f-\lambda]$
``contamination'' from the states $|11 \rangle$ and $|00 \rangle$,
which cannot be arbitrarily minimised without jeopardising the
adiabaticity of the transitions at points $A$ and $C$ (see later).

\begin{figure}[ht]
\begin{center}
\epsfig{file=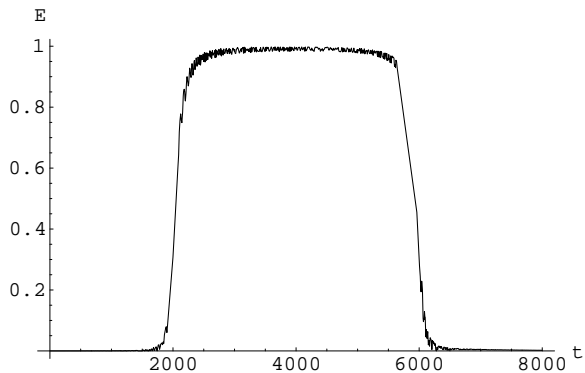,width=8cm} \caption{\label{ent} Entanglement
of $|\phi(t) \rangle$, where $|\phi(0) \rangle = |11 \rangle$, as
a function of time. We take the parameters
$\omega=0.05 \lambda$ and $f(t)=(-2+t/2000)\lambda$, for arbitrary $\lambda$.
The time, t, is given in units of $1/\lambda.$}
\end{center}
\end{figure}

Figure \ref{ent} illustrates the evolution of the entanglement for
$|\phi (t) \rangle$, and we remind the reader again that the Bell
state $|\Psi^- \rangle$ cannot be generated in this manner because
of symmetry requirements.
In this figure, the degree of entanglement is defined to be
the {\it entropy of entanglement} $E$ which is the von Neumann
entropy of the reduced density matrix obtained by tracing out
either of the two qubits \cite{bennett},
\begin{equation}
\label{entanglement}
E = - \text{Tr} \rho_1 \log_2 \rho_1 = - \text{Tr}
\rho_2 \log_2 \rho_2,
\end{equation}
where $\rho_{1} (\rho_{2})$ is the density matrix obtained after tracing out
particle 2 (1).
The related concept, {\it entanglement of formation} $E_f$, to be 
used below, is a
generalisation of the entropy of entanglement for a {\it mixed}
state of two qubits \cite{wootters}.

The entanglement
will stay approximately constant at $E \simeq 1$ when $f$ is in
the range $-\lambda < f < \lambda$. 
Note that if the two qubits had different oscillations frequences, 
$\omega^{(i)}$, mixing between the almost degenerate states
$|\Psi^{\pm} \rangle$ is not prohibited by permutation symmetry
(or the lack thereof). In this case, the degree of entanglement will 
vary significantly in the period $-\lambda < f < \lambda$, attaining
the maximal value of $E \simeq 1$ only at the point $f=0$.

Observe that if we had started instead in the state $|00\rangle$
which is approximately the highest energy eigenstate, we could in
principle obtain the entangled state $\frac{1}{\sqrt{2}}(|11
\rangle + |00\rangle)$ at the point $f=0$. 
However, an inspection of the Hamiltonian
\begin{equation}
H_2^{\rm sym}= \left( \begin{array}{c|c}
                \begin{array}{ccc}
                2f+\lambda & \sqrt{2} \omega & 0  \\
                \sqrt{2} \omega & - \lambda & \sqrt{2} \omega  \\
                0  & \sqrt{2} \omega & -2 f + \lambda   \\
                \end{array}
                & 0 \\
                \tableline
                0 & - \lambda \\ \end{array}
                \right),
\end{equation}
rewritten in the symmetrised basis $\{|11 \rangle, |\Psi^+
\rangle, |00 \rangle, |\Psi^- \rangle\}$, shows that the splitting
between the relevant energy eigenstates for the $|00\rangle
\rightleftharpoons |11 \rangle$ transition at $f=0$ is of order
$\omega^2/\lambda$, as opposed to $\sqrt{2} \omega$ for the $|11
\rangle \rightleftharpoons |\Psi^+ \rangle$ transition considered
earlier. It follows that the resonance width at point $B$ is
necessarily some $\omega/\lambda$ times narrower than those at
points $A$ and $C$.  
Their respective adiabaticity parameters, defined as
\begin{equation}
\gamma \equiv \left| \frac{(\Delta k)^2}{\frac{d}{dt}
\sqrt{(\Delta k)^2 -  (\left.\Delta k\right|_{\rm res})^2 }}
\right|_{\rm res},
\end{equation}
where $\Delta k$ is the splitting between the energy eigenstates,
``res'' denotes evaluated at resonance, and $\gamma >1$ indicates
an adiabatic process, must also differ correspondingly by some
factor of $\omega^2/\lambda^2$.  Explicitly,
\begin{equation}
\gamma_{A,C} \sim \frac{4 \omega^2}{\left|\dot{f} \right|}, \qquad
\gamma_B \sim \frac{2 \omega^4}{\lambda^2} \frac{1}{\left| \dot{f}
\right|}
\end{equation}
characterise the three resonances.  

Hence, in order to manufacture the state 
$\frac{1}{\sqrt{2}}(|11
\rangle + |00\rangle)$, a slowly varying $f$ together with control
over its accuracy to within $\pm \omega^2/\lambda$ are necessary.
Na\"{\i}vely, the severity of the precision requirement seems
easily alleviated by enlarging $\omega$.  The benefit of a wider
resonance width, however, is compensated for by a concomitant
increase in the contamination from the $|\Psi^+ \rangle$ state
which contributes at order $\omega/\lambda$.  By comparison, the
production of $|\Psi^+ \rangle$ from $|11 \rangle$ requires only
that we stop the evolution somewhere between $-\lambda < f <
\lambda$, and is therefore a much simpler task.

\subsubsection{System-Environment coupling: cold bath}

In the presence of environmentally-induced noise, 
we must solve a master equation
for the reduced density matrix of the system.\footnote{Certain
conditions must be satisfied for the dynamics of an open quantum
system to be well-described by a master equation. We consider
only such cases in this paper.} The form of the master equation
depends on whether or not the environment is hot enough to
induce transitions to excited states of the double wells.
We first consider coupling to a cold environment, so that
each particle remains in the ground state of its double well.

\begin{figure}[ht]
\begin{center}
\epsfig{file=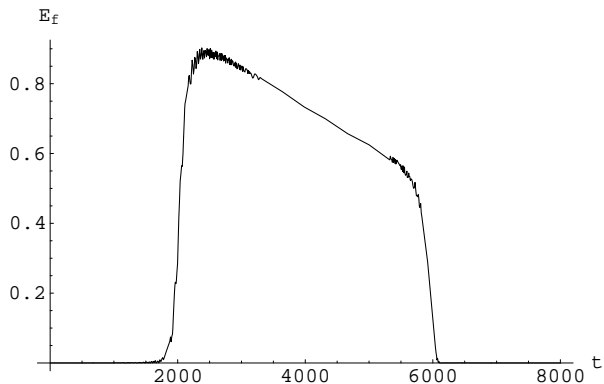,width=8cm} \caption{\label{noise}
Entanglement of formation as a function of time in the presence of noise. 
We take $\Gamma_{\text{relax}}=0.5\times 10^{-4} \lambda$ and display the 
time, $t$, in units of $1/\lambda$.
The 
other parameters the same as in Fig.\ \ref{ent}}
\end{center}
\end{figure}

The evolution of $\rho$ is governed by
\begin{equation}
\label{asnoise}
\frac{d}{dt} \rho(t) = -i [H,\rho(t)]
-\Gamma_{\text{relax}}[\zeta, [\zeta, \rho]]
-\Gamma_{\text{relax}}[\zeta', [\zeta', \rho]],
\end{equation}
where $\zeta$ and $\zeta'$ are Hermitian matrices
obtained from the system-environment interaction Hamiltonian.
For the sake of the example, we will choose
$\zeta = \frac{1}{2}(\identity + \sigma_3 ) \otimes \identity$, 
$\zeta' = \identity \otimes\frac{1}{2}(\identity+ \sigma_3)$, 
which means that the 
relaxation terms arise from a system--bath interaction 
that couples to the left states $| 1 \rangle $ but not to
the right states $| 0 \rangle $.\footnote{The other extreme has
the environment coupling equally strongly to the left and right
sides of the wells, so that $\zeta$ and $\zeta'$ are proportional to the
identity, and the double-commutator terms vanish. We will
return to this point shortly.} 
Commencing with a pure state $|11 \rangle$ and a small relaxation rate
$\Gamma_{\text{relax}}$, we show in Fig.\ \ref{noise} the
evolution of the entanglement with the course of time. 
A ``decay'' in $E$ in the period $-\lambda < f <
\lambda$ is evident. For a sufficiently large
$\Gamma_{\text{relax}}$, the density matrix simply becomes
proportional to the unit matrix, and
the entanglement goes to zero at all times.

\subsubsection{System-Environment coupling: hot bath}

We now consider coupling to an environment hot enough that the system-bath
interactions will cause particles in the wells to populate states
higher than the ground state.
In a previous publication we considered in detail the effects of noise coming 
from a thermal bath coupled to the full excited state spectrum of a 
double well \cite{bsv}. Among the cases
considered was the effect of ``side-blind'' noise on a number of properties
of the system. For the side-blind case, the bath couples {\it equally} to
the left states $|1\rangle$ and the right states $|0\rangle$.\footnote{In the
absence of excitation to higher energy states, side-blind noise does nothing
of note.} If we begin with a single particle on 
the left in the ground state of the symmetrical well, the thermal excitation
of states that have very different R-L oscillation frequencies 
replaces with a multiperiodic tangle the 
previous simple oscillation of the probability of 
finding the particle on the left at a later time.
The effect of a small system-bath interaction rate is to 
gradually damp these oscillations, leaving a time independent 50\% occupancy
on either side. As the ratio of the system-bath interaction rate to
the oscillation rate is increased, however, the motion becomes more organized, 
and in the limiting case where this ratio becomes very 
large,\footnote{
Note that by ``hot bath", we refer to the situation where the
{\it temperature} of the bath is high enough to excite states
other than the ground state.  The {\it coupling}, however, which
determines the {\it rate} of these transitions, is a separate   
parameter.  The ``large $\Gamma$ limit" arises when the
time between successive system-bath interactions is much
smaller than the period of the L-R precession. Observe that
this condition can actually be fulfilled for {\it any} value of
the coupling (and importantly, for a coupling sufficiently
weak that one may treat it perturbatively in the derivation of
the master equation) by making the free precession
period sufficiently large.  In the context of a double well
oscillator this is achieved very easily, simply by
adjusting the central barrier between the wells.  Also
note that this ``large $\Gamma$ limit" is identical to the limit
where one obtains the familiar Quantum Zeno Effect, if the
coupling to the environment is through a $\sigma_3$ operator 
rather than the identity.}
we retrieve sinusoidal oscillation with a frequency that is given 
by the thermal
average of the frequencies of the individual levels \cite{bsv}. The analogous 
phenomenon is called ``motional narrowing" in NMR studies.

We will now look at entanglement-formation dynamics in a two-particle system
with excited states.
The L-R oscillation properties for a particular well-level
are given by Eq.\ (\ref{asnoise}), but the oscillation parameters 
are different for the excited well states. We shall refer to the variable 
distinguishing these states as the ``vertical'' variable, as opposed to the $(L,R)$
degree of freedom which we will term the ``horizontal'' variable. 
We will take the coupling among the vertical levels to again be ``side-blind'' , that is,
not dependent on the horizontal variables, because that is the most interesting case. 
We can capture the essential
features of the problem by considering only two vertical levels, with equal 
transition rates, $\Gamma$, each to the other (the latter as in the case of a high temperature 
for the thermal bath), in which case we make the replacements in Eq.(\ref{asnoise}),
$\rho(t)\rightarrow \rho(E_i,t)$, $i=1,2$  on the lefthand side and
\begin{equation}
\Gamma_{\rm relax}[\zeta,[\zeta,\rho]]\rightarrow  \Gamma \Bigr( \sum_j\rho(E_j,t)-2\rho(E_i,t) \Bigr)
\end{equation}
on the righthand side, as well as specifying, within $H$, different oscillation parameters 
for the two energy states.

Using this equation we now look at the results for the two particle-entanglement
for the cases of zero, moderate, and large system-bath interaction rates.  The 
results are shown 
in Fig.\ 4. We see substantial degradation of entanglement-formation in the case of 
moderate interaction rate. But in the case of large interaction rate, we see a 
pattern of entanglement
creation that is similar to that in the case with no noise. In this limit, the system
is made to change vertical state so rapidly that it can only respond
to an average Hamiltonian, with quantal coherence fully maintained \cite{bsv}.
For these reasons, the entanglement-formation dynamics succeeds in this limit.

\begin{figure}[ht]
\begin{center}
\epsfig{file=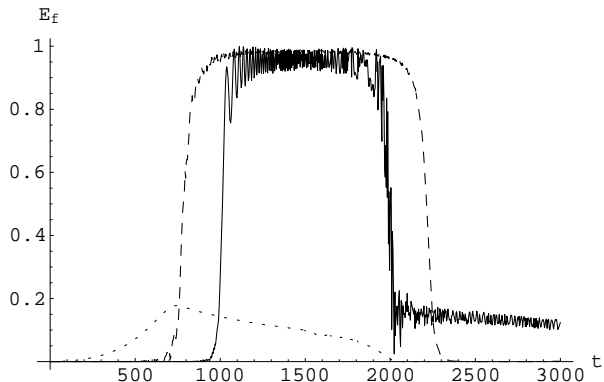,width=8cm} \caption{\label{fbnoise}
Entanglement of formation, $E_f$ as a function of time, in the presence of ``side-blind''
noise.  The solid, dotted and dashed curved are for zero, moderate amd large system-bath
interaction rates, $\Gamma=$ 0, 1 and 1000 respectively.  For simplicity, we took 
a system with only two vertical states, and the initial condition was chosen to be 
$| 11\rangle$, with both particles in their ground states.
We have taken the parameters $\omega(E_0)=0.05$,
$\omega(E_1)=0.1$, 
$\lambda(E_i,E_j)=20 \sqrt{\omega(E_i) \omega(E_j)}$ and 
$f=-3+t/500$, where $\omega$, $f$, $\lambda$ and $\Gamma$ are in units of an
arbitrary energy parameter $E_{{\rm norm}}$, and the 
time in units of $1/E_{{\rm norm}}$.
Note that the position of the level crossing differs for the solid and dashed curves 
as $\lambda$ has been taken to be dependent on the value of the vertical variable.}
\end{center}
\end{figure}

For completeness, we mention that
if the bath does not couple with equal strength to the two horizontal states
$|0\rangle$ and $|1\rangle$, the behaviour is qualitatively the same 
as that obtained from Eq.\ (\ref{asnoise}) for the cold bath, 
where $\zeta \neq \identity$.

\subsection{Multi-Qubit system}


Consider now the $W$ states,
\begin{eqnarray}
|W \rangle_{001} &=& \frac{1}{\sqrt{3}}(|001 \rangle + |010
\rangle + |100 \rangle), \nonumber \\ |W \rangle_{110} &=&
\frac{1}{\sqrt{3}}(|110 \rangle + |101 \rangle + |011 \rangle).
\end{eqnarray}
These may be considered maximally entangled 3-qubit states, in the sense that each 
has the maximum
amount of bipartite entanglement if any one of the three qubits is
traced out \cite{dur}. 

It turns out that the production of a three-qubit $W$ state from
an initially factorisable state such as $|111 \rangle$ via
adiabatic transitions is completely analogous to the generation of
the symmetric Bell state from $|11 \rangle$ considered previously.
We now elaborate on this point.

\begin{figure}[ht]
\begin{center}
\epsfig{file=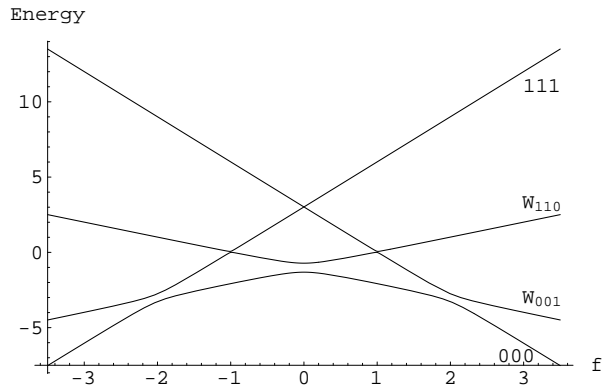,width=8cm} \caption{\label{3qubit} Energy 
level diagram for the set of four permutation symmetric states of the 
three-qubit system, with both the energy and well bias $f$ 
given in units of $\lambda$.  
Note that none of the levels cross.}
\end{center}
\end{figure}

The Hamiltonian for our three-qubit system is a direct
generalisation of that in Eq.\ (\ref{2ham}):
\begin{equation}
H_3= \omega \left[\sigma_{x1} + \sigma_{x2} + \sigma_{x3} \right]
+ f(t)\left[\sigma_{z1} + \sigma_{z1} + \sigma_{z3} \right]  +
\lambda \left[ \sigma_{z1}\sigma_{z2} + \sigma_{z2}\sigma_{z3} +
\sigma_{z3}\sigma_{z1} \right], \label{3ham}
\end{equation}
where, for simplicity, we are dealing explicitly with the
permutation invariant case. Decomposition of the tensor product $2
\otimes 2 \otimes 2 = 4 \oplus 2\oplus 2$ reveals that the
permutation symmetric set $\{|111 \rangle, |W \rangle_{110},|W
\rangle_{001},|000 \rangle\}$ can be treated in isolation, with
its evolution governed by the $4 \times 4$ block in the
Hamiltonian
\begin{equation}
\label{matrixham}
H_3^{\rm sym}= \left( \begin{array}{cccc}
3 f + 3 \lambda   & \sqrt{3} \omega & 0 & 0\\
                             \sqrt{3} \omega & f - \lambda  & 2 \omega & 0 \\
                0 & 2 \omega & -f-\lambda & \sqrt{3} \omega  \\
                0 & 0 & \sqrt{3} \omega & -3 f +3 \lambda \\
                \end{array}\right),
\end{equation}
written in the symmetrised basis.  The level crossing diagram for this set of 
states is plotted in Fig.\ref{3qubit}.

Three resonances mark the evolution of the lowest energy
eigenstate from $f < -2 \lambda $ to $f > 2 \lambda$.  To zeroeth
order in $\omega$, the identity of the state is adiabatically
transformed as per the evolutionary sequence
\begin{equation}
|111 \rangle  \stackrel{f=-2 \lambda}{\longrightarrow} |W
\rangle_{110} \stackrel{f=0}{\longrightarrow} |W \rangle_{001}
\stackrel{f=2 \lambda}{\longrightarrow} |000 \rangle,
\end{equation}
where the conditions above the arrows denote the points at
which the transitions take place.  Commencing with the product
state $|111 \rangle$ at $f < -2 \lambda $, the two $W$ states can
be easily obtained respectively in the domains $-2\lambda < f<0$
and $0<f<2 \lambda$.

Note in passing that the state 
$|{\rm GHZ}\rangle= | 111 \rangle+ | 000 \rangle$ \cite{ghz}
coincides approximately with an
eigenstate at $f=0$ and can in principle be accessed from $|000
\rangle$ or $|111 \rangle$ via the highest energy eigenstate.
However, as a direct analogue of the $\frac{1}{\sqrt{2}}(|11
\rangle + |00 \rangle)$ state considered in the previous section,
the production of the GHZ state by our method suffers the same
problem of non-adiabaticity and again requires  precise control
over the value of $f$.

The generalised form  of the $W$ state for $N$ qubits is defined
as
\begin{equation}
| W \rangle_N = \frac{1}{\sqrt{N}} |N-1,1 \rangle,
\end{equation}
where $|N-1,1 \rangle$ denotes the totally symmetric state with
$N-1$ qubits in the state $| 0 \rangle$ and one qubit in the state
$| 1 \rangle$.  

The residual bipartite entanglement between any two pairs of qubits, 
upon tracing the other qubits out, has been shown to have the maximum
value attainable in an $N$ qubit system \cite{2/n}.
Since $|W \rangle_N$ states are symmetric with respect to
permutation of the $N$ qubits, they can  always be made from $|11
\cdots 1 \rangle$ or $|0 0 \cdots  0 \rangle$ in the manner
discussed above.

\section{Error detection and correction}

An important problem in the study of quantum information theory
is the protection of information from various types of errors.
Any practical realisation of a quantum computer will see the
physical systems, the states of which comprise the qubits, experience effective
violation of unitary evolution and error generation, both due to coupling
to their environment. A number of error detection and protection schemes have been
devised that encode a given qubit through entanglement with other 
otherwise redundant qubits \cite{zurek} \cite{shor}.  Our goal in this section is
to construct a master equation that seamlessly merges
error generation, protection through entanglement and detection within a single
dynamical framework. While the resulting evolution can be usefully understood
through reduction to information theoretic subprocesses, at
the level of pure physics the single equation that one solves encapsulates both
the quantum processes over which the experimenter has no control, and the
knobs the experimentalist turns to change the biases.

In what follows, averted level crossing  schemes similar to those described 
in the previous section 
are used to encode a qubit in a way that it is
protected from certain outside disturbances, following the
steps:
\begin{enumerate}

\item 
We consider an initial pure state given by $|\Psi \rangle=a|11 
\rangle+b|01\rangle \equiv [a|1 \rangle+b|0\rangle] |1\rangle $.  Our goal is 
to preserve the information encoded in the first qubit. 

\item 
We take a noise term that is derived from an external bath that flips 
qubits, independently, for one or both particles, and we follow the density 
matrix for the resulting mixed state.

\item
We transform the state in the time interval $t=0$ to $t=t_e$ by 
adiabatically changing the coefficients of certain terms in the Hamiltonian; 
we refer to these functions as the bias functions.  Next we leave the system 
in this state for a length of time $t_h$ (``hang time'') {\it long} 
compared to $t_e$. 

\item
To reclaim the initial state we adiabatically restore the biases to their 
initial values and then measure the value of the second, control, coordinate. 
If the measurement gives back the original state $|1\rangle$, then for the first 
qubit we obtain the state that is the projection of our evolved density matrix 
on that state. The coding algorithm is such that this operation reclaims 
the density matrix of the initial pure state, with small degradations coming 
from the small non-adiabaticity due to doing the encoding 
and decoding over finite time intervals, or with degradations from letting the 
hang time be so large that multiple errors are likely to have accumulated. If, 
on the other hand, we measure the value $|0\rangle$ for the control state, 
then, beginning from the projection of our evolved density matrix on the 
state $|0 \rangle $ a simple, and implementable unitary operation restores the 
initial density matrix.

\end{enumerate}

We begin this process by describing the system in terms of the four time 
dependent instantaneous eigenstates of the Hamiltonian that describes the 
system without noise, $|\xi_j \rangle_t $, where ${j=1,4}$. Defining , 
$|\xi_j \rangle_0 \equiv|\xi_j \rangle_{t=0}$, we take
initial values,
\begin{eqnarray}
|\xi_1 \rangle_{0}=|1\rangle \otimes |1 \rangle
\nonumber\\
|\xi_2 \rangle_{0}=|0\rangle\otimes | 1\rangle 
\nonumber\\
|\xi_3 \rangle_{0}=|1\rangle\otimes | 0 \rangle
\nonumber\\
|\xi_4 \rangle_{0}=|0\rangle \otimes |0 \rangle 
\label{initiales}
\end{eqnarray}

The initial data is to be encoded in a combination 
of $|\xi_1 \rangle_{0}$, and $|\xi_2 \rangle_{0}$. 
We define a time dependent 
projection operator on the 
subspace used for encoding, 
\begin{equation}
P_e(t) \equiv |\xi_1 \rangle_t\langle \xi_1 |_t
+|\xi_2\rangle_t \langle \xi_2|_t
\end{equation}
We note that $P_0 \equiv P_e(0) = \identity
\otimes|1\rangle\langle 1|$, the projection operator on the 
value unity for the second qubit.
We choose the Hamiltonian
\begin{equation}
H = 2\,f(t)\, \sigma_{x} \otimes \sigma_{x}
+  \sigma_{z} \otimes \sigma_{z}
+ 4*[1 - f(t)]\,  \sigma_{z} \otimes \identity,
\label{newham}
\end{equation}
where $f(t)$ is a function that begins at zero at $t=0$, rises to unity at $t=t_e$, 
remains unity until $t=t_e+t_h$, and then decreases back to zero at  
$t=2t_e+t_h$. At $t=0$, the eigenstates of $H$ are the vectors shown in Eq.(\ref{initiales}). 
We define $U$ to be the time evolution operator for the interval
$t=0$ to $t=t_e$, under the action of this Hamiltonian. If the encoding time is chosen
sufficiently large that the evolution is adiabatic, then for our encoding states we have,
\begin{eqnarray} 
U|\xi_1\rangle_{0}=|\xi_1\rangle_{t_e}=e^{i \phi_1}(|11\rangle + |00\rangle)/\sqrt{2}
\nonumber\\
U|\xi_2\rangle_{0}=|\xi_2\rangle_{t_e}=e^{i \phi_2}(|10\rangle - |01\rangle)/\sqrt{2} 
\end{eqnarray}
where the phases $\phi_1$ and $\phi_2$ are irrelevant to the remainder of the discussion.
These remain eigenstates until $t=t_e+t_h$. 

We will show that this dynamics protects against ``bit-flip'' errors defined by choosing the
coefficients $\zeta$, $\zeta'$, etc. in an equation like \ref{asnoise} to be from the list of
error operators $\sigma_x \otimes \identity$ and $\identity  \otimes
\sigma_x$, plus ``phase errors'' induced by
$\identity \otimes \sigma_z$ and $\sigma_z \otimes \identity$.
(It is easy to modify the scheme to protect against all
four bit-flip errors $\sigma_{x,y} \otimes \identity$ and
$\identity \otimes \sigma_{x,y}$, but at the cost of
losing the phase error protection.)
The key to this protection  
is that the any error operator $\zeta$, picked from the above set
has all matrix elements equal to zero in the subspace spanned by 
$|\xi_1\rangle _{t_e}$ and $|\xi_2 \rangle_{t_e}$. 
\footnote{There will of course be nonvanishing matrix 
elements from this subspace to the subspace spanned by 
$|\xi_3 \rangle_{t_e}$ and $|\xi_4 \rangle_{t_e}$. }
We can express this property using the projection operator, $P_e(t)$,
\begin{equation}
\zeta=P_e(t)\,\zeta [1-P_e(t)]+[1-P_e(t)]\zeta \,P_e(t)+[1-P_e(t)]\zeta [1-P_e(t)]
\label{proj2}
\end{equation}
which holds throughout the time interval $t_e<t <t_e+t_h$, noting that $P_e(t)$ is independent
of time during this interval.

In what follows, for clarity of exposition, we shall leave the noise turned off during the
short 
encoding-decoding times;
however in the numerical example presented later, we retain the noise throughout 
the process.
We begin with a density matrix,
\begin{equation}
\rho(0) = \left[a |\xi_1\rangle_0 + 
b |\xi_2\rangle_0\right]
\left[ a^* \langle \xi_1| _0+ b^* 
\langle \xi_2|_0\right],
\end{equation}
and let it evolve in a unitary fashion to $t=t_e$,
\begin{equation}
\rho(t_e)= U^{-1}\, \rho(0)\, U
=  \left[a |\xi_1\rangle_{t_e} + 
b |\xi_2\rangle _{t_e}\right]
\left[ a^* \langle \xi_1|_{t_e} + b^* 
\langle \xi_2|_{t_e}\right] + \text{n.a.}
\end{equation}
where ``n.a.'' stands for small 
terms coming from non-adiabaticity, which can 
induce hopping into the $|\xi_3\rangle$ and $|\xi_4\rangle$ states.
Between $t=t_e$ and $t=t_e+t_h$ the time evolution is governed by the 
equation,
\begin{equation}
\frac{d}{dt} \rho(t) = -i [H(t_e),\rho(t)]
-\Gamma [\zeta\, ,\, [\zeta\, ,\, \rho(t)]]
= -i [H(t_e),\rho(t)] -\Gamma \left\{\rho(t) - \zeta \rho(t) \zeta \right\},
\label{withnoise}
\end{equation}
where the noise term is of the same form as given in
Eq.\ (\ref{asnoise}), the constant 
$\Gamma$ sets the level of the noise, 
and $\zeta$ is one of the error matrices enumerated above, with $\zeta^2=1$. 
We shall be concerned with the changes to the density matrix that are first order
in $\Gamma$. However we note that if we omit the last term in Eq.(\ref{withnoise}), the 
 solution, $\rho_1$, would be just a damped form of the solution without noise, which is just
the evolution under the Hamiltonian, $H(t)$, constant in the interval $\,t_e<t <t_e+t_h\,$,
\begin{equation}
\rho_1(t_e + t) =  P_e(t_e)\,\exp[-i H(t_e) t]\, U^{-1}\, \rho(0)\, U\, 
\exp[i H(t_e) t]\, P_e(t_e) \, \exp[-\Gamma \, t].
\label{damped}
\end{equation}
The right and left projection operators are unnecessary here, since nothing has 
evicted us from the encoding subspace,
but they are useful for the next step.
For the effects on $\rho (t_e + t_h)$ of the $\zeta$ dependent term in Eq.(\ref{withnoise}), 
which we denote as $\rho_2$,  we can calculate the change in 
$\rho (t_e + t_h)$
to first order in $\Gamma$ by direct time integration after inserting the ($\Gamma=0$) 
form of  $\rho(t)$ from Eq.(\ref{damped}) into
the last term on the RHS of Eq.(\ref{withnoise}). For our purposes all we need to know about this
term comes
from the projection operator structures in Eq.(\ref{proj2}) and Eq.(\ref{damped}) which directly
give,

\begin{equation}
\rho_2 (t_e + t_h) =  [1 - P_e(t_e+t_h)]\, \rho_A \, [1 - P_e(t_e+t_h)] + O(\Gamma^2)
\label{afternoise}
\end{equation}
where $\rho_A$ is the result of the time integral discussed above. The total correction is
given by
$\rho_1+\rho_2$.

Next we decode, by adiabatically changing the parameters back to their 
values at $t = 0$. This is
a unitary process implemented by the time evolution operator $\overline U$
for the interval $t = t_e + t_h$ to $t = 2t_e + t_h$. We note that 
$\overline U P_e(t_e+t_h) \overline U^{-1} = P_0$,
since the unitary evolution processes never cause a transition from one of the 
states
$|\xi_j \rangle _{t}$ to another in the adiabatic limit, and the relative phase factors 
from the time evolution cancel
in the construction of the projection operators. Thus we obtain,

\begin{eqnarray}
\rho(2t_e+t_h) = &P_0\, \overline U\, \exp(-iHt_h)\, U^{-1}\, \rho(0)\, 
U\, \exp(iHt_h)\, \overline U^{-1}\, (1-\Gamma t_h)\, P_0
\nonumber\\
&+(1 - P_0)\, \overline U\, \rho_A\, \overline U^{-1}\,
(1 - P_0) + O(\Gamma^2)
\label{finalrho}
\end{eqnarray}
In the $|\xi_1 \rangle_0$, $|\xi_2 \rangle_0$ 
subspace the unitary matrix $\overline U\, \exp(-iHt_h)\, U^{-1}$ is diagonal. 
Thus if we measure the second (control) variable 
and obtain unity (or no change from the initial value) there can have been no 
bitflip in the state of the first (information) state. Then, if we wish we can 
fine-tune the hang time, $t_h$, so that 
$\overline U\, \exp(-iHt_h)\, U^{-1}$ in the 
$|\xi_1\rangle_0$, $|\xi_2\rangle_0$ subspace
is a phase factor times the unit matrix. In this case we have 
recaptured the initial state of the information bit perfectly, using 
implementable transformations that are independent of the information itself.

The above process can be characterized as being a limited error correction, 
in that we have corrected the errors that could have been created in the part of 
the final density matrix in which the value of the control bit is unity. In the 
case that the measurement of the control bit gives zero (this probability is 
itself of order $\Gamma t_h$), then we can work out the explicit form for 
the second term on the righthand side of Eq.\ (\ref{finalrho}), 
and find a unitary operation 
that converts
it to a multiple of the initial density matrix. 
This operation depends on which of the four error 
operators is used for the matrix $\zeta$. Thus if we
extended Eq.\ (\ref{withnoise}) to include several error terms
at once, then, since we would not know which was operative
in a given case, we could not implement the correction.
Corrections of all of the error operators require 
encoding protocols involving more qubits, as discussed 
in Ref.\ \cite{shor}, and in much subsequent literature.
        
However, if we had replaced the double-commutator on the
righthand side of Eq.\ (\ref{withnoise}) by a sum of terms, one for each
of our protectable error operators, the solution would still
have the useful property of error detection. In this case
we can view the result in Eq.\ (\ref{finalrho}) as preservation
for the part where the control bit is unity, and 
with error detection in the cases in which measurement
of the control bit gives zero. When we measure zero for the 
control bit, we must discard the information in the channel. 
But even a single redundant parallel channel 
with the same input data would reduce the probability of information loss to 
the level of order $(\Gamma t_h)^2$. Since the latter is the essential level 
of error for the case in which measurement of the control bit gives unity,
this is the best result obtainable using our kind of mechanism.

\begin{figure}[ht]
\begin{center}
\epsfig{file=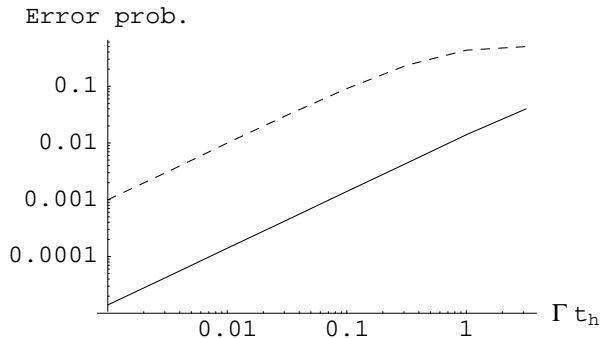,width=8cm} \caption{\label{error}
The solid curve gives the bitflip error probability as a function of
$\Gamma t_h$ for the case in which the encoding and decoding
mechanisms are in place. The encoding time was taken to be
$t_e = 0.01\times t_h$. The dashed curve gives the error generated
with the same noise source but without the encoding.}
\end{center}
\end{figure}

In our numerical example, we simply solve Eq.\ (\ref{withnoise}) for the 
density matrix evolution in the presence of noise, 
beginning at time $t=0$ and ending at time $t=2t_e+t_h$, without breaking 
into the separate factors of the discussion above. This then includes every 
piece of Eq.\ (\ref{finalrho}), 
but with the additional corrections due to: a) Small 
non-adiabaticity (hopping) due to
the fact that time taken by the encoding process is not infinite. b) Errors that 
occur from the noise interaction during encoding-decoding, due to the fact 
that the encoding is not instantaneous. c) The terms of higher order in 
$\Gamma t_h$, which matter when the probability of more than one error 
during the time $t_h$ becomes appreciable. For our example, we choose the 
initial state for the information bit to be simply $|1\rangle$. 
Figure \ref{error} shows the probability of a bit 
flip error arising from an error operator $\zeta=\sigma_x \otimes \identity$ 
plotted as a function of the average number of applications of the error, 
$\Gamma t_h$, and on the same plot the results for the case in which the 
information is not encoded. We have also checked numerically to see that in 
cases in which we begin with a superposition of $|1\rangle$ and $|0\rangle$ 
for the information bit, and fine tune $t_d$ to an appropriate value, the entire 
state, including the relative phase angle in the superposition, is preserved to 
the same degree as the bit-error values plotted in Fig.\ \ref{error}.

\section{Conclusion}

We have considered two related topics in the study of closed and open quantum
system evolution that go beyond one-body system Hamiltonians by including
interaction terms between the system particles. In the first topic,
we have shown that entangled states that are symmetric
under the permutation of the basis states can be generated via adiabatic
transitions from initially factorisable states which have this
symmetry.  In a two-qubit system, the outcome is the symmetrical Bell
state $\frac{1}{\sqrt{2}}(|10\rangle+|01\rangle)$, while the three-qubit
analogue is the set of two $W$ states. Noise in moderate quantitites 
generally reduces
entanglement over time, and for such cases the limit of large noise sees 
entanglement remaining zero at all times. However, for the special
case of ``side-blind'' noise that leads to motional narrowing, the
creation of entanglement is preserved despite a large system-environment
interaction rate.

In our second topic, we have shown that for small noise levels, 
we can use an entangling-pause-unentangling cycle to store and 
recover quantum data in a two qubit system, 
with error probability quadratic rather than linear in the parameter $\Gamma T$ where
$\Gamma$ is the error rate for unencoded data and $T$ is the time elapsed.
Calculations of encoding that utilise the entanglement
give a concrete demonstration of the compatibility of two requirements on
the encoding time: a) that it be long enough achieve adiabaticity;
b) that it be short enough to limit the effects of noise during the
encoding process.

\end{document}